# SUB-NANOMETER HEIGHT SENSITIVITY BY PHASE SHIFTING INTERFERENCE MICROSCOPY UNDER AMBIENT ENVIRONMENTAL FLUCTUATIONS

Azeem Ahmad[1,2,a], Vishesh Dubey[1,2], Ankit Butola[1], Jean-Claude Tinguely[2], Balpreet Singh Ahluwalia[2,] and Dalip Singh Mehta[1,b]

[1]*Department of Physics, Indian Institute of Technology Delhi, Hauz Khas, New Delhi 110016, India*

[2]*Department of Physics and Technology, UiT The Arctic University of Norway, Tromsø 9037, Norway*
*Corresponding authors: [a]ahmadazeem870@gmail.com
Email: balpreet.singh.ahluwalia@uit.no & [b]mehtads@physics.iitd.ac.in

**Abstract**

Phase shifting interferometric (PSI) techniques are among the most sensitive phase measurement methods. Owing to its high sensitivity, any minute phase change caused due to environmental instability results into, inaccurate phase measurement. Consequently, a well calibrated piezo electric transducer (PZT) and highly-stable environment is mandatory for measuring accurate phase map using PSI implementation. Here, we present a new method of recording temporal phase shifted interferograms and a numerical algorithm, which can retrieve phase maps of the samples with negligible errors under the ambient environmental fluctuations. The method is implemented by recording a video of continuous temporally phase shifted interferograms and phase shifts were calculated between all the data frames using newly developed algorithm with a high accuracy $\leq 5.5 \times 10^{-4}$ π rad. To demonstrate the robustness of the proposed method, a manual translation of the stage was employed to introduce continuous temporal phase shift between data frames. The developed algorithm is first verified by performing quantitative phase imaging of optical waveguide and red blood cells using uncalibrated PZT under the influence of vibrations/air turbulence and compared with the well calibrated PZT results. Furthermore, we demonstrated the potential of the proposed approach by acquiring the quantitative phase imaging of an optical waveguide with a rib height of only 2 nm. By using 12-bit CMOS camera the height of shallow rib waveguide is measured with a height sensitivity of 4 Å without using PZT and in presence of environmental fluctuations.

## 1. Introduction

Digital holographic microscopy (DHM) is a quantitative phase imaging (QPI) technique that has the capability to record the complex wavefront (amplitude as well as phase) of the light field interrogated with the specimen [1-3]. In digital interference microscopy, the complex information about the specimen is encoded in the form of spatially modulated signal generated due to its coherent superposition with the reference field and eventually captured by the area detector, i.e., charge coupled device (CCD)/complementary metal oxide semiconductor (CMOS). The encoded information can be further recovered by employing either single-shot or multi-shot phase retrieval algorithms [4, 5] depending upon the recording geometries broadly classified as an off-axis and an on-axis interferometric configurations[6-8].

An off-axis digital interference microscopy can recover information related to specimen from a single interferogram, which makes it suitable to study dynamical behaviour of biological cells or tissues [3, 6, 8]. However, it utilizes maximum one-fourth of the detector's bandwidth for the noise (DC and twin image) free phase recovery. As a consequence, resolution of the reconstructed object field is limited irrespective of the detector's capability, which records diffraction limited information about the specimen. To overcome this limitation, on-axis digital interference microscopy attracted strong attention of many researchers, which can utilize full resolving power of CCD/CMOS cameras [9]. Nevertheless, it needs multiple phase shifted interferograms for noise (DC and twin image) free phase recovery of biological specimens at full detector resolution [9].

To date, a significant amount of work has been done in the development of various phase shifting interferometry (PSI) techniques[10]. PSI techniques have been most commonly used for various applications in optical metrology, digital holography and quantitative phase microscopy [1, 7, 11]. However, most of the PSI techniques utilize expensive piezo electric transducer (PZT) to introduce equal amount of phase shift ($\delta$) between sequentially recorded interferograms [7, 11-13]. Further, PSI is extremely sensitive to rapid and imprecise movement of PZT and environmental instability because fringe formation results from the path differences of the order of wavelength of light. Even a small amount of error in PZT calibration, which occurs due to the backlash and hysteresis, can lead to the nonlinear phase shift between data frames resulting in phase measurement errors, such as background modulation errors in the reconstructed phase maps of the target, see results of Refs. [5, 14-21].

The background modulation errors have been minimized previously by employing various generalized phase shifting algorithms given in Refs. [5, 21-25]. However, with these algorithms, the modulated phase error has not been completely removed in the reconstructed phase maps. In addition, most of the algorithms are implemented only for the simulated and industrial objects such as low roughness gauge block and resolution chart. The performance of previous algorithms [5, 21-25] in terms of phase sensitivity has not been tested on the objects with small phase shifts such as small optical thickness of sub-10 nm. Further, PSI based on such algorithms would also require expensive phase shifting devices like PZT, spatial light modulators, digital micro mirror devices etc. and sophisticated electronic control systems to introduce precise phase shift between data frames. In addition, the major source of error confronted in PSI is external vibrations, such as air turbulence and it has been reported that this greatly affects phase shift and subsequently phase measurement accuracy and precision [10, 26]. To overcome these problems, various single-shot or simultaneous PSI techniques have been developed in the past [27-30]. These techniques require either four CCD camera with complex experimental setup or single CCD camera having a micro-polarizer array to record four phase shifted interferograms simultaneously. This either increases cost of the system significantly or leads to inefficient utilization of CCD chip as each four phase shifted interferogram uses only one fourth pixels of the CCD chip [27, 28]. All these aforementioned issues related to PSI compel us to think about a straightforward and cost effective method which can avoid the use of expensive phase shifting devices, large number of the optical components and most importantly alleviate the problem of external air turbulence and vibrations.

In this paper, we propose a novel approach that is capable to retrieve phase information of the specimen with good accuracy and high precision without the need of PZT calibrated phase shifted frames. The method is implemented with the recording of a time lapsed interferometric movie of the specimens in presence of external vibration/air turbulence. Subsequently, accurate measurement of the phase shift between the individual interferometric frames of the recorded movie is done using the developed algorithm. This eradicates the use of any expensive phase shifting devices for the implementation of PSI techniques to perform phase/height measurements of the specimens. The method can even utilize external air turbulence/vibrations to introduce phase shift between frames to determine the phase maps related to specimens. Various unconventional methods such as cell phone vibration, artificial air turbulence generated from hair dryer, and manual translation of reference or sample arm was utilized to introduce temporal phase shift in the data frames and consequently utilized for successful phase reconstruction.

First, the influence of two important parameters phase shift error ($\alpha_t$) during recording of phase shifted interferograms and utilization of wrong phase shift value ($\delta$) during reconstruction on the phase measurements of specimens is studied both through simulations and experiments. It is observed that precise knowledge of these two parameters is mandatory for the accurate assessment of phase maps using phase shifting based DHM/QPI techniques.

The present approach is then tested on a standard optical strip waveguide structure and compared with the results obtained from PZT assisted PSI. Further, it is implemented for the QPI of human red blood cells (RBCs) and shallow rib waveguide structures (2 – 8 nm). The detector's noise is found to be the most limiting factor for QPI of the objects having shallow optical height like rib waveguide. A systematic study is done to understand the effect of detector's noise on QPI of 8 nm rib waveguide by employing three scientific cameras with different specifications. Among three different cameras, CMOS sensor having 12-bit depth is found to be suitable for accurate quantitative phase measurement of 2 – 8 nm rib waveguides (phase sensitivity of <4 Å). Another, advantage of the proposed approach is that it can be implemented for any number of step, for example, 3-step, 4-step,………, N-step PSI depending upon the speed of recording device. This is advantageous as the phase measurement accuracy increases with the increase in phase steps [14, 31]. In the present study, we will discuss advantages, results and opportunities associated with our present method.

## The proposed method: Theory and simulations

### 2.1 Five frame phase shifting algorithm

The five frame phase shifting algorithm is widely preferred over the other phase shifting interferometry because of moderate phase error and acquisition time [5]. There is a trade-off between phase shift error and acquisition time, i.e., if one tries to decrease the phase shift error by increasing the number of frames (say N), the acquisition time gets increased or vice versa. However, phase shift error cannot be completely removed even for N-step phase shifting algorithm because of the environmental instability. Hariharan[5] proposed a five frame phase shifting algorithm for the phase measurement with acceptable phase measurement error. The intensity modulation of five phase shifted 2D interferograms at a particular wavelength recorded by WL-PSIM can be expressed as follows[5]:

$$I_1(x,y) = A(x,y) + B(x,y) + 2\sqrt{A(x,y)B(x,y)} \cos[\varphi(x,y) - 2\delta + \alpha_t] \qquad (1)$$

$$I_2(x,y) = A(x,y) + B(x,y) + 2\sqrt{A(x,y)B(x,y)} \cos[\varphi(x,y) - \delta] \qquad (2)$$

$$I_3(x,y) = A(x,y) + B(x,y) + 2\sqrt{A(x,y)B(x,y)} \cos[\varphi(x,y)] \qquad (3)$$

$$I_4(x,y) = A(x,y) + B(x,y) + 2\sqrt{A(x,y)B(x,y)} \cos[\varphi(x,y) + \delta] \qquad (4)$$

$$I_5(x,y) = A(x,y) + B(x,y) + 2\sqrt{A(x,y)B(x,y)} \cos[\varphi(x,y) + 2\delta] \qquad (5)$$

where, A$(x,y)$ and B$(x,y)$ are the intensities of two light beams. $\varphi(x,y)$ is the phase information related to test object, δ is the phase shift between two consecutive phase shifted frames and $\alpha_t$ corresponds to the time varying phase shift error between data frames due to external vibrations. The phase information '$\varphi(x,y)$' related to test object can be calculated from the following expression if $\alpha_t = 0$[5]:

$$\varphi(x,y) = \tan^{-1}\left[\sin\delta\frac{2(I_4(x,y) - I_2(x,y))}{I_1(x,y) - 2I_3(x,y) + I_5(x,y)}\right] \quad (6)$$

The phase map can be further utilized to calculate corresponding height map '$H(x,y)$' using the following relation:

$$H(x,y) = \frac{\lambda}{4\pi(n_s(x,y) - n_m(x,y))}\varphi(x,y) \quad (7)$$

where, $\lambda$ is the central wavelength of light source, $n_s(x,y)$ is the refractive index of sample and $n_m(x,y)$ is the refractive index of surrounding medium.

A simulation study is done to understand the effect of both the phase shift 'δ' between two frames (as described in Eq. 6) and the time varying phase shift error '$\alpha_t$' between the frames (as described in Eq. 1) on the quantitative phase measurement that occurs during the recording of phase shifted interferograms. Phase shifting interferometry (PSI) or holography techniques are very sensitive to external vibrations or air turbulence, which makes them challenging to implement for accurate quantitative imaging of specimen. In addition, an uncalibrated phase shifter can introduce an unwanted error in the phase measurement. The collective effect of errors of the phase measurement is simulated and studied for five frame phase shifting algorithm [5].

## 2.2. Simulations

### 2.2.1 Influence of '$\alpha_t$' and 'δ' on the phase measurement

First, the effect of '$\alpha_t$,' on the phase measurement during the reconstruction step using Eq. 1 is studied. Five equal phase shifted interferograms are simulated, four of them having equal phase shift ~ $70^0$ and the other one having different phase shifts (determined by $\alpha_t$), further details can be found in Supplementary Fig. S1. These five phase shifted frames are utilised to retrieve phase map using Eq. 6. To understand the effect of '$\alpha_t$', a range of values from – $20^0$ to $20^0$ at an interval of $5^0$ are used in Eq. 1. Note $\alpha_t = 0^0$ corresponds to no phase shift error. The value of 'δ' equal to $70^0$ is used into Eq. 6 during reconstruction. It is observed from the simulation results that phase shift error '$\alpha_t$' leads to the generation of modulated background noise in the reconstructed phase maps (see Fig. 1a). In addition, it is observed that the frequency of the modulated background noise is twice than that of the interferograms spatial frequency as illustrated in Supplementary Fig. S2. This exhibits a good agreement with the previous studies [5]. The reconstructed phase maps corresponding to each value of phase shift error are shown in Fig. 1a. Importantly, it can be clearly visualised from the phase images that peak to valley (PV) phase error gets prominent as the value of '$\alpha_t$' is deviated from true phase shift.

Figure 1(b) depicts the line profiles of the reconstructed phase maps corresponding to a range of phase shift error '$\alpha_t$' used in the Eq. 1. Note that background modulation follows the symmetric trend as '$\alpha_t$' go away from $0^0$ towards lower and upper side. Figure 1(c) illustrates the plot of maximum PV error as a function of '$\alpha_t$'. The peak to valley phase error is found to be equal to 200 mrad for $\alpha_t$ = -20⁰ and 20⁰, whereas, zero exactly at $\alpha_t$ = 0⁰. Therefore, in order to overcome such type of severe issues from the phase images, phase shift error must be equal to zero during the recording of all phase shifted interferograms. This situation is analogous to the nonlinear drift of the phase shifter, which could arises due to the hysteresis and thermal drift of PZT [5]. In addition, air turbulence or vibration could also affect the measurement and introduce unwanted phase shift error '$\alpha_t$' between the frames. This can also introduce a significant error in the phase measurement if the non-linearity is large. The nonlinearity can introduce unequal phase shift only in one or more than one phase shifted frame keeping equal phase shift between rests of the frames. Even a small phase drift in only one of the frame will results into rather large reconstruction artefacts as seen in Fig. 1.

Next, effect of '$\delta$' on the phase measurement during reconstruction step using Eq. 6 is studied. The phase shift error '$\alpha_t$' is kept zero for this study. This is analogous to the situation in which prior information about the equal phase shift between data frames is unknown. It is observed that use of wrong '$\delta$' rather than the actual phase shift value adopted by phase shifter may also introduce background modulation error during reconstruction as shown in Fig. 1d. In order to realize it numerically, a range of phase shift '$\delta$' from 50⁰ to 90⁰ in a step of 5⁰ is introduced in Eq. 6. The reconstructed phase maps corresponding to above range of '$\delta$' are presented in Fig. 1d. Figure 1e depicts line profiles corresponding to all the reconstructed phase maps. Figure 1f illustrates the variation of maximum PV error as a function $\delta$. It is evident from Figs. 1c and 1f, that the variations of PV error as a function of '$\alpha_t$' and '$\delta$' are not the same. It is noteworthy that phase shift error '$\alpha_t$' and/or insertion of wrong value of '$\delta$' in Eq. 6 during reconstruction leads to the generation of serious modulated background phase error in the phase map. The phase error will even increases further if both the values ('$\alpha_t$' and '$\delta$') are wrong during the reconstruction.

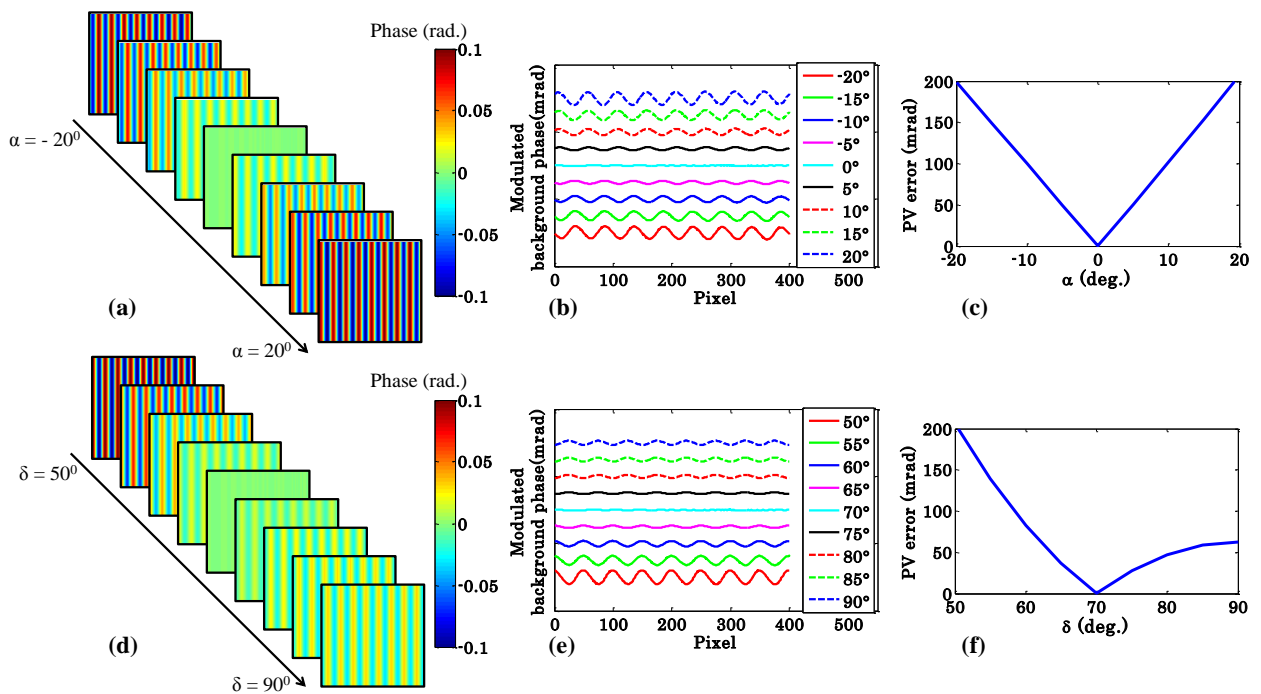

**Figure 1** Simulation results to understand the effect of phase shift '$\delta$' and phase shift error '$\alpha$' on the phase measurement. (a) Reconstructed phase map from the simulated five equal phase shifted plane

wave interferograms except one, which have a range of phase shift error '$\alpha_t$' from $-20^0$ to $20^0$ at an interval of $5^0$. (b) Corresponding line profiles along white dotted line of reconstructed phase maps. (c) PV phase error as a function of phase shift '$\alpha_t$'. (d) Reconstructed phase map from the simulated five phase shifted (=$70^0$) plane wave interferograms for a range of '$\delta$' from $50^0$ to $90^0$ at an interval of $5^0$, which are inserted into Eq. 6 during reconstruction. (e) Corresponding line profiles along white dotted line of reconstructed phase maps. (f) PV phase error as a function of '$\delta$'.

### 2.2.2 Proposed phase shift measurement algorithm

Any phase shift '$\delta$' during reconstruction and/or any phase shift error '$\alpha_t$' during recording significantly affects the phase measurements as seen in Fig. 1. For an accurate phase measurement it is necessary to effectively suppress the background phase error from the reconstructed phase maps. The conventional way of performing this is to use costly opto-mechanical components (such as SLM, PZT etc.) to ensure that well calibrated, pre-determined and equal phase shifts between the interferograms can be generated. Here, we take an inverse approach where we allow phase to drift with time and record a continuous movie followed by a computational method which can extract five (or more) equally phase shifted interferograms for retrieval of the phase information. Here, a time lapsed interferometric movie of the test object, which contains all the phase shifted interferograms (equal and unequal both), is recorded. The interferometric movie is further drag and drop into ImageJ software to extract all the frames of the movie. The unknown phase shifts between all the extracted frames make the accurate phase extraction difficult. We developed an algorithm to calculate the phase shift between frames accurately, which enables the use of five frame phase shifting algorithm for accurate phase retrieval related to object.

To calculate the phase shift '$\delta$' between phase shifted interferograms, following steps were adopted:

(1) Draw a line profile along a line perpendicular to the interference fringe pattern, where no object structure is present,
(2) To increase the measurement accuracy, the line profile is interpolated using spline interpolation,
(3) Sinusoidal fit the data obtained from step (2), which further increases the accuracy of the phase shift measurement,
(4) Calculate the difference '$\Delta$' between two consecutive maxima of a sinusoidal fitted function acquired from step (3), and multiply the inverse of $\Delta$ with $2\pi$ to calculate the pixel value of interferogram in terms of degree, i.e., 1 pixel = $2\pi/\Delta$.
(5) Repeat steps 1 – 3 for all the frames extracted out from the movie, and calculate the difference '$\Delta_1$' between maxima of sinusoidal fitted function of the first frame and rest of the frames.
(6) Finally, phase shifts of all the frames with respect to the first frame are calculated by the product of $\Delta_1$ and pixel value obtained at step (3), i.e., phase shift can be calculated by using the following expression: $\delta = \Delta_1 (2\pi/\Delta)$.

The diagrammatical illustration of the sequence of steps above is presented in Fig. 2.

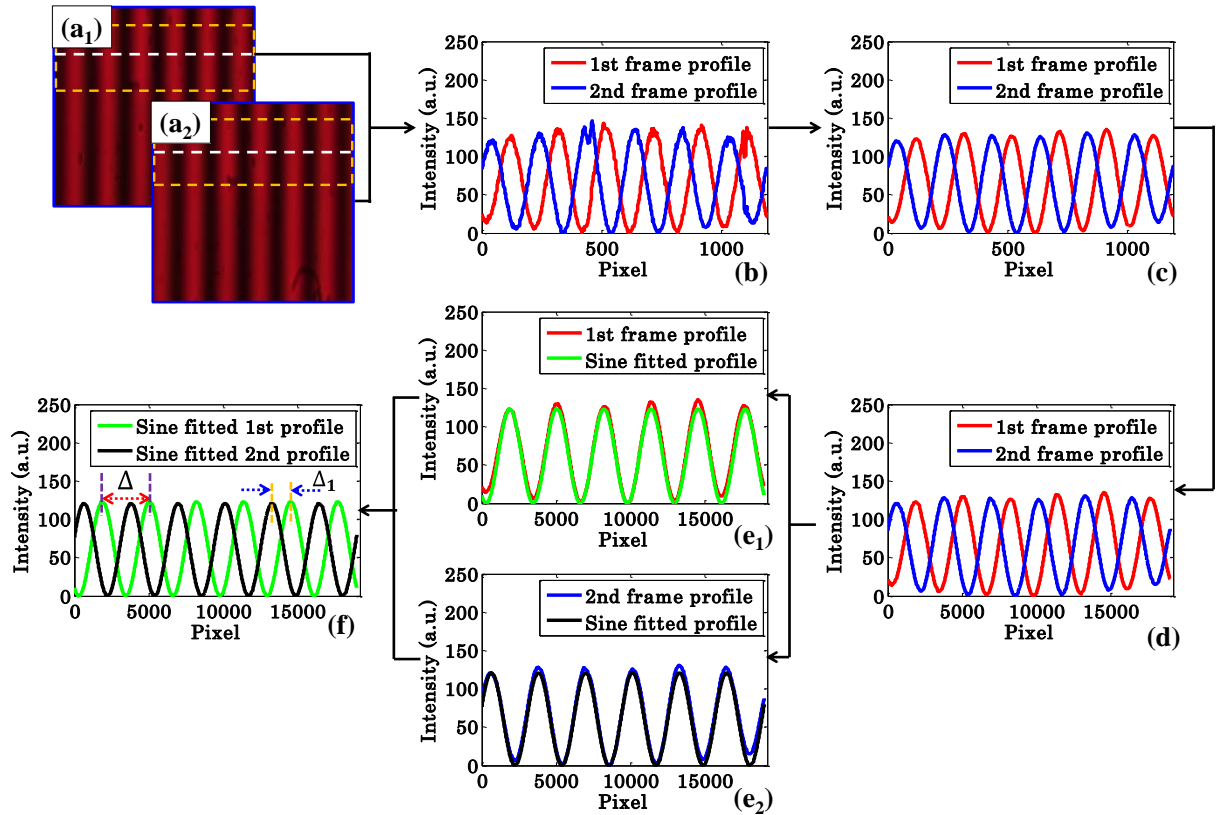

**Figure 2** Diagrammatical illustration of the sequence of steps followed in developed phase shift measurement algorithm.

Figure 2a depicts two different interferograms having some unknown phase shift between them. To calculate phase shift between them line profiles were taken along the direction perpendicular to the fringe, i.e., white line for both interferograms shown in Fig. 2a. The corresponding line profiles are presented in Fig. 2b. It can be seen that line plots do not have smooth profiles, which make difficult to locate peak positions of fringe patterns. To increase the accuracy of measurement, average of all the line plots along rows coming in the dashed yellow box is taken. Figure 2c shows averaged line profiles of 1st and 2nd frame. It is evident from Fig. 2c, averaged line profiles are smoother than the line profiles drawn along single row, which increased the accuracy to find the peaks of interferometric signals. To further improve the accuracy of our method, both the line profiles are spline interpolated to get intermediate data points (Fig. 2d). Finally, sinusoidal fitting of both the interpolated line profiles are done to calculate phase shifts with accuracy ~ $5.5 \times 10^{-4}$ π rad. between interferograms as shown in Fig. 2e. The phase shift measurement accuracy strongly depends upon the sampling interval of the data during interpolation. Therefore, finer sampling of the data during interpolation can further increase the phase shift measurement accuracy. The value of $\Delta_1$ shown in Fig. 2f provides amount of phase shift between two interferograms and can be calculated by following steps 5 and 6 of the proposed algorithm.

## 3. MATERIALS AND METHODS

### 3.1 Experimental details:

To realize the effect of equal and unequal phase shifted interferograms on the phase measurement experimentally, we employed white light phase shifting interference microscopy (WL-PSIM) as depicted in Fig. 3. Further, same optical setup is utilized to exhibit the capability of the proposed algorithm incorporated with five frame phase shifting algorithm for accurate phase measurement of industrial and biological cells.

The experimental scheme of the present setup is based on the principle of non-common-path white light interference microscopy. The narrow bandpass filter having peak wavelength 620nm with ~ 40nm spectral bandwidth is inserted into white light beam path for the recording spectrally filtered interferogram. The image of the light source is relayed at aperture stop (A-stop) plane using lens $L_1$, where size of aperture controls spatial coherence of the light source. The source image is further relayed at the back focal plane of the objective lens with the help of lens $L_2$ and beam splitter BS to achieve uniform illumination at the sample plane. The field stop (F-stop) controls the field of view (FOV) of the microscope. The beam splitter BS directs the beam towards Mirau interferometric objective lens (50×/0.55 DI, WD 3.4 Nikon, Japan) to generate white light interferograms easily (inset of Fig. 3). More details about the Mirau interferometric objective lens can be found elsewhere [13, 32]. The Mirau interferometer is attached with PZT to introduce required temporal phase shift between data frames. The phase shifted data frames are then captured using CCD/CMOS camera for further analysis.

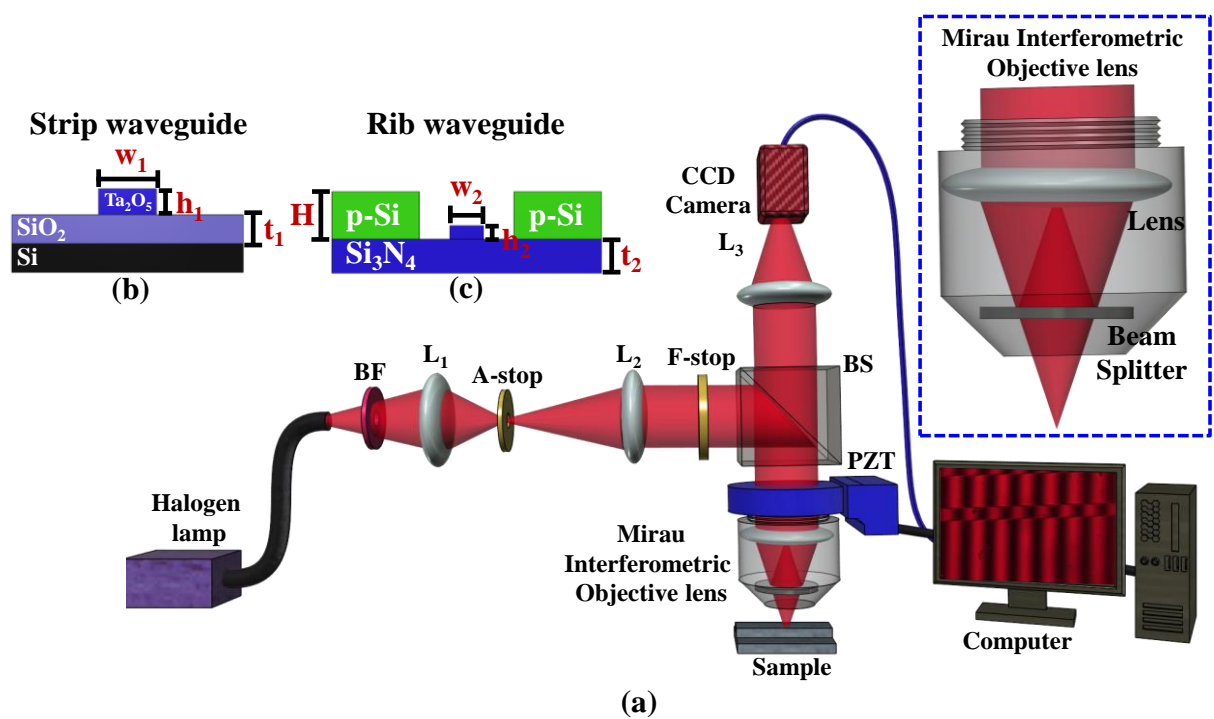

**Figure 3** (a) Schematic diagram of WL-PSIM setup. $L_1$, $L_2$ and $L_3$: lenses; BS: beam splitter; A-stop: aperture stop; F-stop: field stop; BF: bandpass filter; PZT: piezo electric transducer; and CCD: charge coupled device. Fig. 1 b-c shows the schematic diagram of two optical waveguide geometries (b) strip, and (c) rib waveguides. Strip waveguide is made of $Ta_2O_5$: tantalum pentoxide (core), $SiO_2$: silicon di-oxide, Si: silicon substrate. Rib waveguide is made of $Si_3N_4$: silicon nitride (core), and p-Si: poly-silicon (absorbing layer). The waveguide parameters are $w_1$, $w_2$ = widths, $h_1$, $h_2$ = strip and rib height, $t_1$ = thickness of $SiO_2$, $t_2$ = slab region of rib waveguide and H = total thickness of absorbing layer.

### 3.2. Test sample

Two different optical waveguides and human blood are utilized as test specimens in the proposed work. The schematic diagrams of the optical waveguide geometries: strip and rib waveguide, are depicted in Figs. 3b-c, respectively. Strip and rib waveguides were fabricated by sputtering a guiding layer of $Ta_2O_5$ [33] and $Si_3N_4$ [34] onto a silica (Si) substrate followed by photolithography and argon ion-beam milling, respectively [35]. For strip waveguides, the layer of $Ta_2O_5$ ($n_{core}$ = 2.12 @ 620nm) had a thickness '$h_1$' of 220 nm and width '$w_1$' of 10μm, and was completely etched down to the $SiO_2$ layer (n = 1.45). For the rib waveguides, the layer of $Si_3N_4$ (refractive index ~ 2.041 at 620nm) is only partially etched down by thickness '$h_2$' leaving a final slab thickness of '$t_2$'. The rib region had a thickness '$h_2$' of 8 nm and width '$w_2$' of 2 μm. The rib waveguide had an absorbing layer (H = 180 – 200 nm) of poly-silicon (p-Si) onto $Si_3N_4$ as illustrated in Fig. 3c. More details on the optimization of waveguide fabrication can be found elsewhere [33, 34]. Strip waveguide geometry is utilized to realize the effect of phase shift error '$\alpha_t$' and phase shift between data frames 'δ' on the phase measurement. Same waveguide geometry is further used to compare the reconstructed phase maps obtained from manual phase shifting and well calibrated PZT. The rib waveguide geometry is employed to investigate the influence of detector noise on the system's phase measurement sensitivity.

For the preparation of biological sample, fresh blood sample is collected from a healthy donor with skin puncture. The fingertip is, first, cleaned with 70% isopropyl alcohol before the skin puncture. The skin is then punctured with one quick stroke to achieve a good flow of blood from the fingertip. The first blood drop is wiped away to avoid the excessive tissue fluid or debris. The surrounding tissues are gently pressed until another blood drop appears. The blood drop is then put onto the reflecting silicon substrate and spread using a glass slide to form a thin blood smear for interferometric recording.

## 4. Experimental results and discussion

### 4.1 Experimental realization of the influence of '$\alpha_t$' and 'δ' on the phase measurement

To experimentally realize the effect of '$\alpha_t$' and 'δ' on the phase measurement, experiments are conducted on an optical waveguide with stripe geometry. The strip waveguide was made of $Ta_2O_5$, of 220 nm thickness and 10 μm wide, others details are presented in section 3.2. The strip waveguide is placed under the WL-PSIM having 620 nm central wavelength bandpass filter (bandwidth ~ 40nm) into the white light beam path to record a time lapsed movie of (frame rate ~15 fps) of the five phase shifted interferograms using uncalibrated PZT (see Visualization 1). For more details about the experimental setup see material and methods section. The interferometric movie of the waveguide is recorded without vibration isolation table under environmental fluctuation to capture phase shifted interferograms. The phase shifted interferograms were then extracted into frames using *ImageJ* software. The phase shifts between first and rest of the frames were, further, calculated by using the developed algorithm. Once the phase shifts between frames are measured, the following studies are performed experimentally: (1) the effect of phase shift 'δ' used at the time reconstruction and (2) effect of phase shift error '$\alpha_t$' which occurs due to PZT uncalibration or air turbulence/vibration.

The five equal phase shifted interferograms were selected from the recorded interferometric movie for the waveguide phase measurement. The frame numbers of the five phase shifted interferograms were found to be 1, 49, 84, 129, 170 frames of the movie having phase shift between consecutive frames approximately equal to 70.8$^0$ using the developed algorithm. Further, these frames are utilized in Eq. 6 to measure the phase map of strip waveguide. Figure 4a depicts the reconstructed waveguide phase map having minimal background noise may be due to the detector's non-linear noise. The magnified view and corresponding line profile of background phase map (region enclosed by the red box) are shown in Fig. 4b. It can be visualized from Fig. 4b; PV phase shift error noise has been completely removed from the phase images with some detector noise ~30 mrad. The studies related to the detector's noise are presented in section 4.4.

To envisage the influence of wrong 'δ' on the phase measurement of waveguide, the value of δ = 52° is put into Eq. 6. The same frames given above are utilized for the reconstruction. The PV value of background modulation is much smaller than the maximum phase value of waveguide. Therefore, δ = 52° is chosen to clearly visualize the background modulation with waveguide phase map as illustrated in Fig. 4c. The reconstructed phase map of waveguide when δ = 52° instead of actual phase shift (i.e., 70.8°) is used in Eq. 6 is shown in Fig. 4c. The background modulation can be clearly visualized in the reconstructed waveguide phase map as depicted in the magnified view of Fig. 4c. Figure 4d shows the line profile along black dotted line in the enlarged region of Fig. 4c as marked with the red box. The PV error generated in the reconstructed phase map is found to be equal to 200 mrad at δ = 52$^0$ (Fig. 4d), which is in a close agreement with the simulation results.

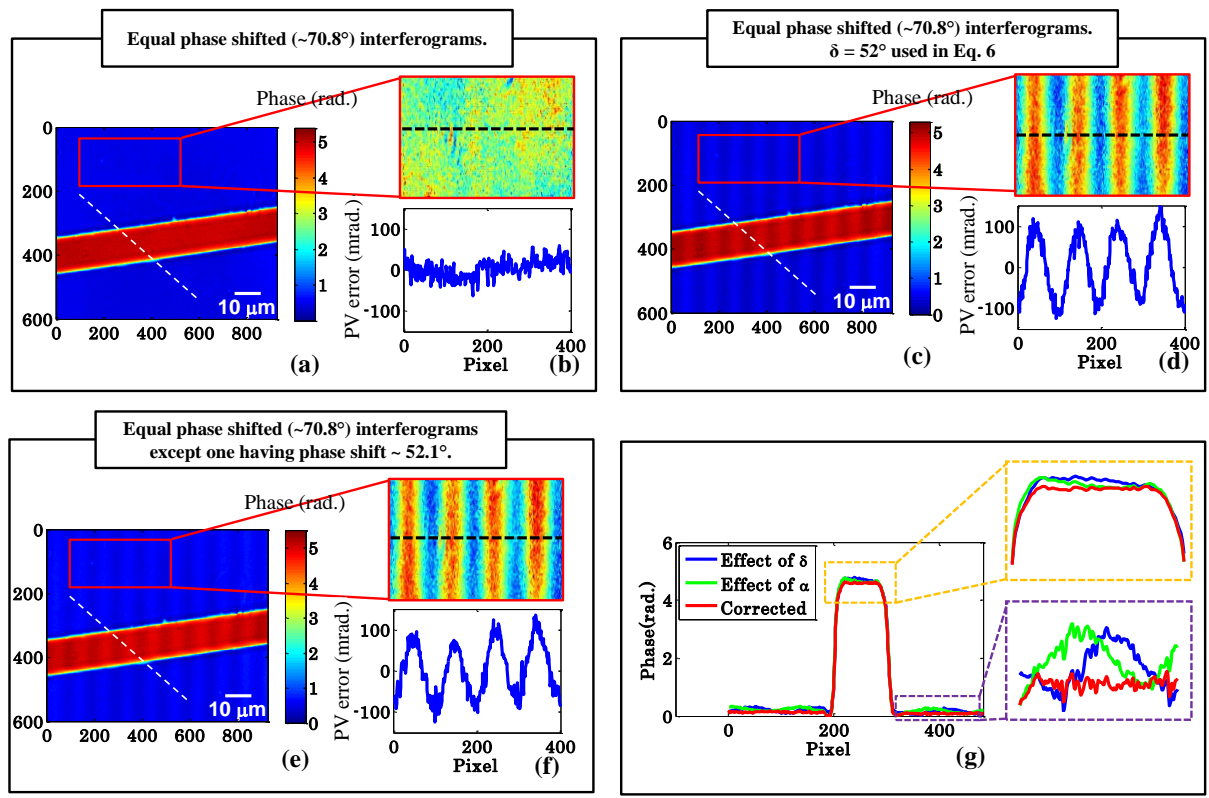

**Figure 4** Experimental investigation of the effect of phase shift 'δ' and phase shift error '$\alpha_t$' on the quantitative measurement of optical waveguide. The test sample is stripe waveguide of 220 nm height. (a) Reconstructed waveguide phase map when all five equal phase shifted (~70.8°) interferograms

obtained from the proposed algorithm are utilized for the reconstruction. (b) Line profile along black dotted line shown in the inset of (a). (c) Reconstructed waveguide phase map when the value of 'δ' equal to 52° instead of 70.8° is utilized during reconstruction using Eq. 6. (d) Line profile along black dotted line shown in the inset of (c). (e) Reconstructed waveguide phase map when all five equal phase shifted (~70.8°) interferograms except one which have a phase shift value equal to 52.1° instead of 70.8° are utilized for the reconstruction. The value of 'δ' equal to 70.8° is used in Eq. 6 for the reconstruction. (f) Line profile along black dotted line shown in the inset of (e). (g) Line profile along white dotted lines shown in (a, c, e). The scale bar is depicted in white color line.

Similarly, to investigate the effect of phase shift error 'α' on the reconstructed phase map, one of the interferometric frame say 49th (phase shift ~ 70.8°) is being replaced from 77th frame (phase shift ~ 52.1°). Figure 4e illustrates 2D reconstructed phase map of the waveguide, when one of the interferogram having phase shift equal to 52.1° is used keeping other phase shifted frames same. This situation is analogous to simulated ones, where unequal phase shift is intentionally introduced into one of the phase shifted interferogram (Fig. 1b). The unequal phase shift between above two frames (1st and 77th) is originated because of the external vibration, hysteresis and nonlinearity of PZT. The insets of Figs. 4c and 4e are the magnified images of background phase map marked with red boxes. Figure 4f presents line profile along the black dotted line. It is illustrated from Fig. 4f; sinusoidal background phase error is generated while using one unequal and rest of the equal phase shifted interferograms in five frame phase shifting algorithm [in Eq. 6]. It is quite evident from Figs. 4d and 4f, PV errors in both the cases are found to be equal to ~ 200 mrad, which exhibit a good agreement with the simulated results. It is therefore necessary to have equal phase steps (i.e. $\alpha_t$ =0) and accurate determination of phase shift between frames (i.e., precise δ).

To further compare noise level of the corrected phase map with the reconstructed ones under the influence of 'δ' and '$\alpha_t$', line profiles were plotted along white dotted lines shown in Figs. 4a, 4c, 4e. It can be envisaged from Fig. 4g, wrong 'δ' during reconstruction and phase shift error '$\alpha_t$' during recording greatly influence the phase measurement of test object. The insets of Fig. 4g depict this in the magnified views of line profiles marked with dotted yellow and purple boxes.

**4.2 A comparison of proposed method and PZT assisted QPI of optical waveguide**

To demonstrate the robustness of the proposed method instead of using piezo-translation stages two simple methods are used to induce the phase shift. First a simple manual translation of reference or sample arm and second by using hair dryer to introduce random phase shift. Both the methods worked equally well but in the present work we included results obtained from manual phase shifting technique. The recorded interferometric movies of the temporal phase shifted frames introduced by hair dryer (Visualization 2) and translation of reference or sample arm manually (Visualization 3) can be found in supplementary note. Temporal phase shift introduced due to the translation of sample arm manually is utilized for quantitative imaging of a stripe optical waveguide and human red blood cells (RBCs).

To demonstrate the phase measurement accuracy while using manual phase shifting technique, results are compared to that of well calibrated PZT. Figures 5a and 5b are the results obtained from well calibrated PZT of a stripe optical waveguide (H = 220 nm). One of the phase shifted interferogram recorded by employing WL-PSIM and calibrated PZT is illustrated in Fig. 5a. Rest of the equal phase shifted frames (phase shift ~ 75°) are shown in Supplementary Fig. S3. The equal five phase shifted frames are further utilized for the reconstruction of phase map of strip waveguide as depicted in Fig. 5b.

To verify the capabilities of proposed algorithm, experiments are performed on the same waveguide. A continuous temporal phase shift in the interference pattern is introduced by translating sample stage manually. The time lapsed interferometric movie of waveguide is then recorded using Infinity2-1RC color CCD camera having frame rate ~ 15 fps (see Visualization 3). Using the proposed methodology (as discussed in Section 3.2) the measurement of phase shift between data frames was determined. The five equal phase shifted (phase shift ~ $69^0$) interferometric frames are selected from the recorded movie as shown in Fig. 5c. These extracted five phase shifted frames are then utilized to calculate wrapped phase map using Eq. 6. To determine unwrapped phase map related to waveguide, a minimum $L^p$ norm 2D phase unwrapping algorithm is applied[36]. The 2D reconstructed phase map corresponding to the region enclosed by red box is presented in Figs. 5e. To compare the reconstructed phase maps obtained from manual phase shifting and well calibrated PZT, line profiles along the white dotted lines shown in Figs. 5b (calibrated PZT) and 5e (manual) are drawn as presented in Fig. 5f. It is worth noting from Fig. 5f that phase maps correspond to calibrated PZT and manual phase shifting are found to be in close agreement with each other. The noise generated in the reconstructed phase maps is due to the camera noise.

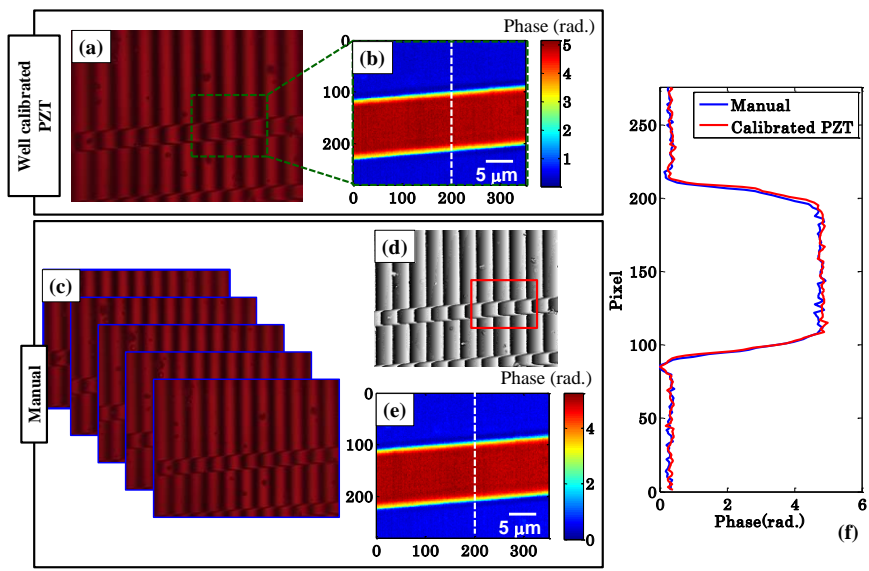

**Figure5** Quantitative phase measurement of optical waveguide using calibrated PZT and unconventionally obtained temporal phase shifted interferograms, which is done by the translation of sample stage manually. The test sample is stripe waveguide of 220 nm height. (a) One of the five equal phase shifted interferogram of optical waveguide obtained from calibrated PZT. (b) Corresponding unwrapped phase map. (c) Five equal phase shifted interferograms of waveguide extracted from a recorded movie using proposed algorithm. (d) Corresponding wrapped phase map. (e) Unwrapped phase map of the region marked with red box. (f) Line profiles along white dotted vertical lines illustrated in (b) and (e). The scale bar is depicted in white color.

### 4.3 Quantitative phase imaging of biological cells

To exhibit potential of the proposed method in the field of biomedical imaging, quantitative phase imaging of human RBCs is done. The details about the RBC sample preparation can be found in section 2.3. A time lapse interferometric movie of human RBCs can be seen in

Visualization 4. The continuous phase shift between data frames is introduced manually. The recorded movie is extracted into frames using ImageJ software. Then developed algorithm is adopted to identify equal five phase shifted interferograms/frames of RBCs. Figure 6a shows five phase shifted interferograms of human RBCs. The wrapped phase map of RBCs is further calculated using five frame phase shifting algorithm Eq. 6 and illustrated in Fig. 6b. The phase ambiguities are removed by employing minimum $L^p$ norm 2D phase unwrapping algorithm [36]. The reconstructed phase map of RBCs enclosed in the inset of Fig. 6b marked with a red box, is depicted in Fig. 6c. It can be seen that 2D phase map of human RBCs has flat background, i.e., no sinusoidal phase error present in the reconstruction. Figure 6d presents line profile of one of the reconstructed RBC along a black dotted line shown in Fig. 6c. The reconstructed 2D phase map and line profile of RBC have donut shape as expected from the previous studies [6].

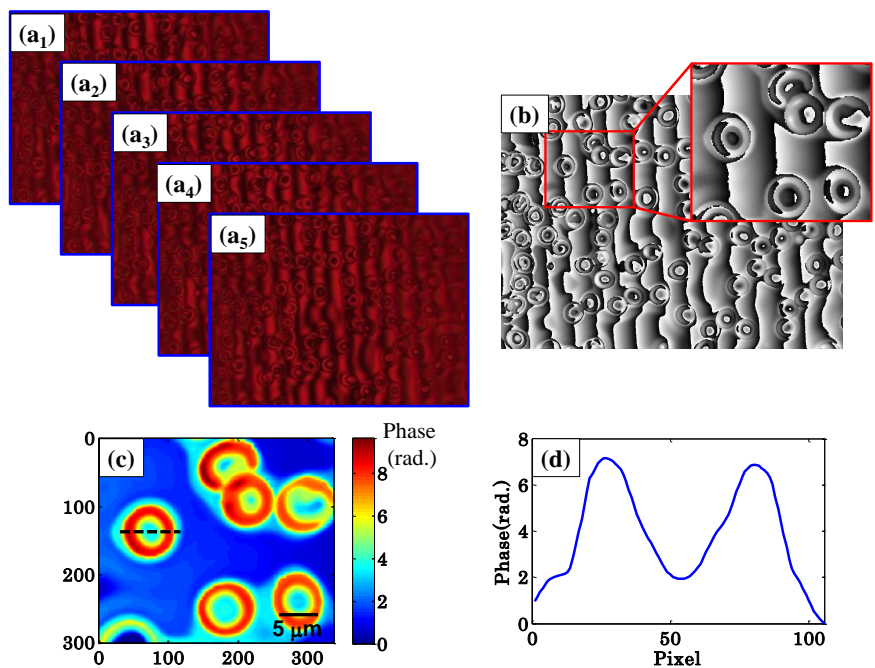

**Figure 6** Quantitative phase measurement of human RBCs using unconventionally obtained temporal phase shifted interferograms, which is done by translating the sample stage manually. (a) Five equal phase shifted interferograms of RBCs extracted from a recorded movie using proposed algorithm. (b) Corresponding wrapped phase map. (c) Unwrapped phase map of the region marked with red box. (d) Line profile of a single RBC along black dotted horizontal line illustrated in (c). The scale bar is depicted in black color.

### 4.4 QPI of sub nanometer optical path length and investigation of detector's noise on overall phase sensitivity

To exhibit the potential of the present technique in terms of optical path length (OPL) sensitivity, experiments are performed on a rib waveguide with rib height of only 8 nm. The small rib height of 8 nm generates very small optical height ~ 8.32 nm at 620 nm wavelength. Additional details of the rib waveguide can be found in section 3.2. Due to small OPL, such samples are difficult to realize under conventional DHM due to coherent noise of laser. Further, conventional WL-PSIM for the phase measurement suffers from the issues of hysteresis and nonlinearity of PZT, and environmental instability. For the successful implementation of the proposed algorithm for OPL measurement sensitivity, influence of detector's noise on the phase/height measurement of 8 nm rib waveguide is systematically studied.

To investigate the influence of detector noise on the system's phase measurement sensitivity, three different cameras: Lumenera color CCD (Infinity2-1RC), Lumenera monochrome CCD (Infinity2-1RM) and Hamamatsu ORCA-spark CMOS cameras (C11440-36U) are employed. Additional details about the cameras, which may play a crucial role in QPI, can be found in the Supplementary Note (see Table S1). The rib waveguide (optical height ∼ 8.32nm) is placed under the WL-PSIM having 620 nm central wavelength bandpass filter (bandwidth ∼ 40nm) into the white light beam path to record temporally phase shifted interferograms manually at ∼15fps (see Visualization 5 and 6). Here, a 15 s interferometric movie is recorded to acquire 200-250 data frames using Infinity2-1RC (color) and Infinity2-1RM (monochrome) camera, which are adequate to obtain five equal phase shifted interferograms for the phase reconstruction. Whereas, only 6s interferometric movie at 34fps (see Visualization 7) is sufficient to provide ∼ 200 data frames while employing 12-bit CMOS camera.

The interferometric movie contains phase shifted interferograms, which are extracted into individual frames using ImageJ software. The phase shifts between first and rest of the frames are then calculated by using the developed algorithm. Once the phase shifts between frames are measured, the equal five phase shifted frames are selected for the calculation of the phase map of waveguide by employing Eq. 6. The details about the five equal phase shifted frames for all three cameras can be found in the Supplementary Note. One of the five phase shifted interferograms are shown in Figs. $7a_1$, $7a_2$ and $7a_3$ corresponding to Infinity2-1RC, Infinity2-1RM and C11440-36U camera, respectively. The extracted data frames are further utilized to measure the phase/height map of 8 nm rib waveguide. The recovered phase maps are illustrated in Figs. $7b_1$, $7b_2$ and $7b_3$. It can be visualized from the retrieved phase map that the background modulation is not observed which is otherwise quite common problem in PSI systems. Further, the recovered phase map images show narrow rib waveguide (of 8 nm rib height) along the middle line of blue color region (Figs. $7b_1$, $7b_2$ and $7b_3$), which is otherwise difficult to visualize in the interferometric images. Figures $7c_1$, $7c_2$ and $7c_2$ illustrate the magnified view of the regions of Figs. $7b_1$, $7b_2$ and $7b_2$ marked with green dotted boxes having 8nm rib height. The enlarged phase maps are further used to measure corresponding height maps using Eq. 7. The height maps of rib waveguide obtained from Infinity2-1RC, Infinity2-1RM and C11440-36U camera are shown in Figs. $7d_1$ $7d_2$ and $7d_3$, respectively. The line profiles of retrieved phase and height maps along white dotted lines are depicted in Figs. $7e_1$ and $7f_1$ corresponding to Infinity2-1RC. Similarly, for Infinity2-1RM and C11440-36U camera line profiles of retrieved phase and height maps along white dotted lines are depicted in Figs. $7e_2$, $7f_2$ and $7e_3$, $7f_3$, respectively.

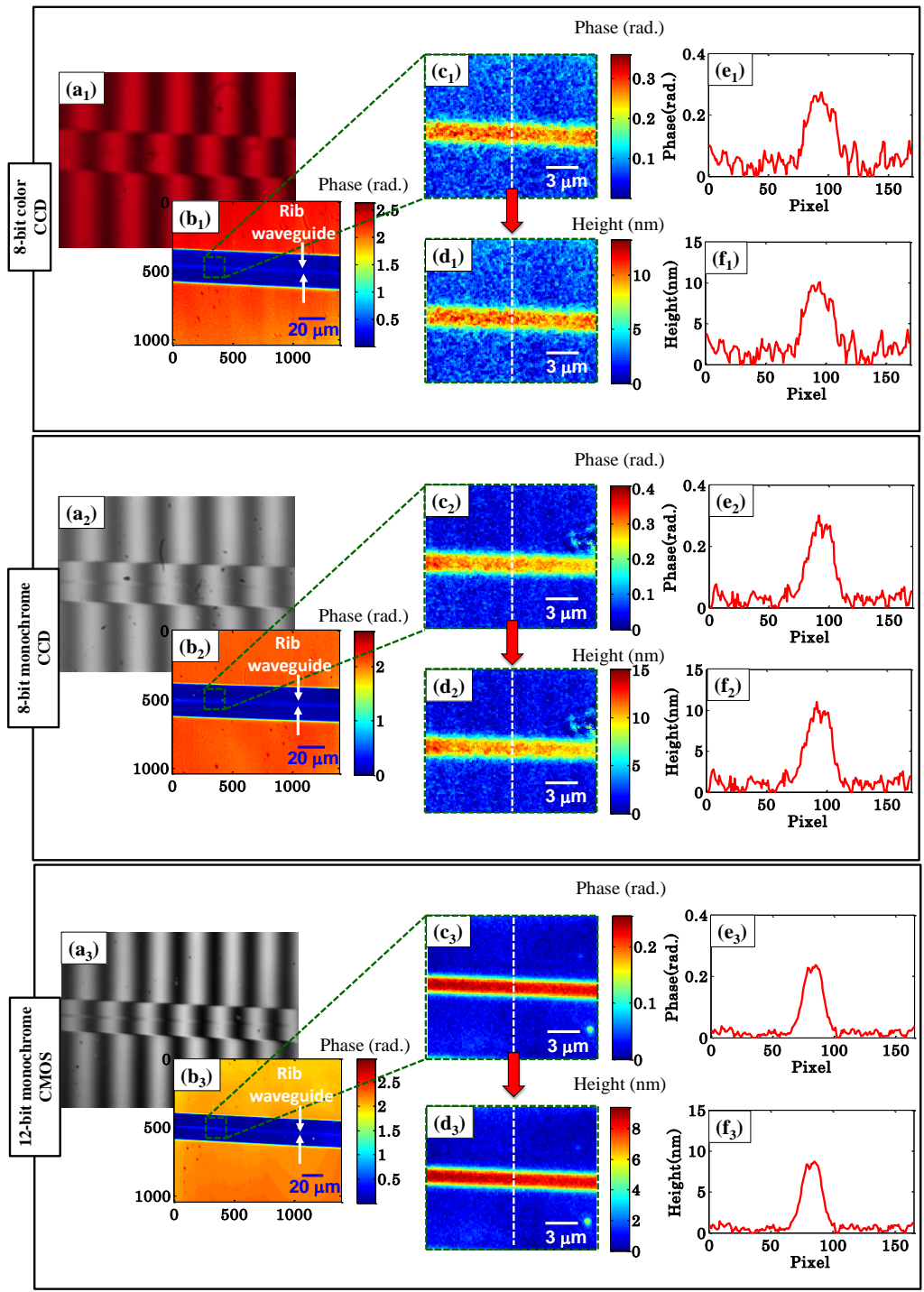

**Figure 7** Quantitative phase/height measurement sensitivity of the present method while using 8-bit color, monochrome CCD and 12-bit monochrome CMOS camera. The test sample is shallow rib waveguide of rib height of only 8 nm. Five temporally phase shifted interferograms are extracted using ImageJ from a movie recorded by the translation of sample stage manually (see visualizations 5 – 7). ($a_1$ – $a_3$) One of the phase shifted interferograms of rib waveguide extracted from a movie recorded by 8-bit color, monochrome CCD and 12-bit monochrome CMOS camera, respectively, using proposed algorithm. ($b_1$ – $b_3$) Corresponding unwrapped phase maps. ($c_1$ – $c_3$ and $d_1$ – $d_3$) Unwrapped phase and height map of the region marked with green dotted box in ($b_1$ – $b_3$). ($e_1$ – $e_3$ and $f_1$ – $f_3$) corresponding line profiles of 8nm rib waveguide along white dotted vertical lines. The scale bars are depicted in blue and white colors.

It can be clearly visualized from the phase line profiles illustrated in Figs. 7e$_1$ and 7e$_2$ that use of color CCD camera generates more noise in the phase reconstruction than monochrome CCD camera. The generation of more phase noise present in the phase reconstruction while using color CCD camera compared to monochrome camera could be due to the color crosstalk of the camera. It is worth noting that the phase/height measurement sensitivity of WL-PSIM system while employing 12-bit CMOS camera is further increased as shown in Fig. 7e$_3$. This could be due to some other properties of the detectors such as bit depth, dynamic range, full well capacity, and read noise. For additional details of 12-bit CMOS camera and its comparison with other cameras, please see Supplementary Table S1[37]. This will also be true for the other existing QPM techniques.

The peak to valley (PV), mean and standard deviation (SD) of the phase noise generated while using three cameras: Infinity2-1RC, Infinity2-1RM and C11440-36U, is presented in Table 1. The height of the rib waveguide is measured to be equal to ∼ 9.6 ± 1.1 nm for color CCD camera. Similarly, for monochrome camera the height of rib waveguide is observed to be equal to 9.4 ± 0.8 nm. Both the measurements are not found to be in a good agreement with actual height of the waveguide (∼ 8 nm). Whereas, the height of the rib waveguide is measured to be equal to ∼ 8.1 ± 0.4 nm, which is in a close agreement with actual value ∼ 7.5 ± 0.2 nm obtained from atomic force microscopy (Bruker Dimension icon), see Fig. 8. The 3D height map of the same rib waveguide is illustrated in Fig. 8a. The line profile of the height map along the green dotted line is depicted in Fig. 8b. The improvement in the phase/height measurement sensitivity of the system while employing 12-bit CMOS camera is due to large bit depth, large dynamic range, large full well capacity, and less read noise compared to 8-bit CCD camera (Please see Supplementary Table S1). The slight inaccuracy ∼ 0.5 nm in the height measurement could be arising due to the following possible reasons: (1) small variations of height for rib waveguide of rib height of 8 nm, (2) detector's noise, and (3) measurement inaccuracy of AFM (∼ 0.6 nm). The measurement error due to detector's noise can be further corrected with the implementation of less noisy 16-bit or cooled CCD/CMOS camera for interferometric recordings, which can be more explored in future. It is worth noting that detector's noise greatly influence the height measurement accuracy of the test specimens especially for the objects having small optical thickness.

**Table 1.** Peak to Valley (PV), Mean, and Standard Deviation (SD) of phase/height measurement sensitivity of the proposed approach for three different cameras: Lumenera color CCD (Infinity2-1RC), Lumenera monochrome CCD (Infinity2-1RM) and Hamamatsu ORCA-spark CMOS cameras (C11440-36U).

| Camera | Phase sensitivity (mrad.) | | | Height sensitivity (nm) | | |
|---|---|---|---|---|---|---|
| | PV | Mean | SD | PV | Mean | SD |
| Infinity2-1RC | 102.70 | 43.55 | 26.22 | 4.52 | 1.91 | 1.15 |
| Infinity2-1RM | 69.10 | 28.44 | 18.11 | 3.04 | 1.25 | 0.79 |
| C11440-36U | 37.40 | 13.14 | 8.26 | 1.64 | 0.63 | 0.40 |

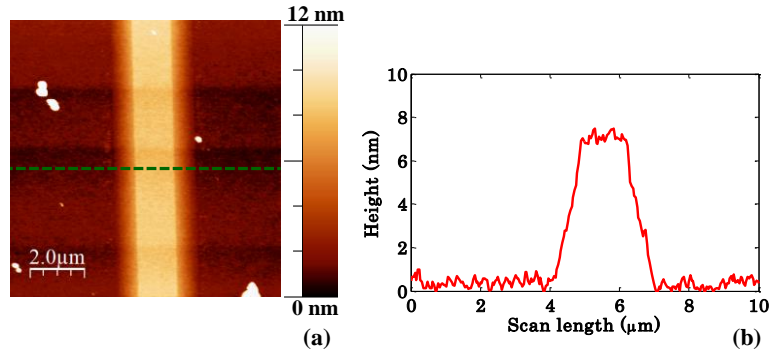

**Figure 8** Profilometry of 8nm optical rib waveguide using Atomic Force Microscopy. (a) Height map of rib waveguide. (b) Corresponding line profile along the green dotted horizontal line.

Next, experiment was performed on optical waveguide with rib height of only 2 nm and width of 100 μm (Fig. 9a). For this experiment 12-bit CMOS camera was used to investigate the accuracy of the proposed method. The movie containing temporally phase shifted interferograms of waveguide with rib height of only 2 nm can be found in supplementary Note (see Visualization 8). The five frames which have equal phase shift (~72°) between consecutive frames are found to be 1, 24, 32, 49, and 71 frames of the movie. One of the five equal phase shifted interferogram is shown in Fig. 9b. The extracted data frames are further utilized to measure the height map of 2 nm rib waveguide. The inset illustrates the magnified view of the region marked with red dotted box in Fig. 9b having 2 nm rib height lies between two blue color dotted horizontal lines. The segmented region is further used to measure corresponding height map using Eq. 7 as depicted in Fig. 9c. The line profile of retrieved height map along black dotted line is depicted in Fig. 9d. The height of the rib waveguide is measured to be equal to ~ 2.3 ± 0.3 nm, which is in a close agreement with the actual value ~ 2 nm. The slight inaccuracy could be again due to the detector's noise.

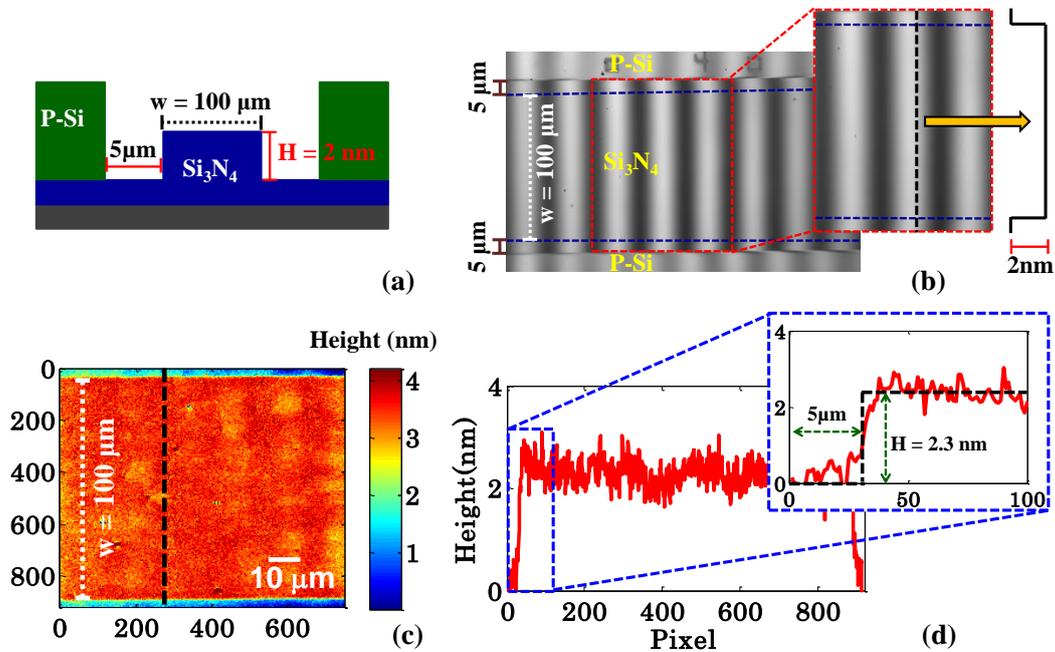

**Figure 9** Height measurement of 2nm rib waveguide while using 12-bit monochrome CMOS camera. (a) Rib waveguide structure. (b) One of the phase shifted interferogram of rib waveguide. (c) Corresponding

height map. (d) The line profile of 2nm rib waveguide along black dotted vertical line. The scale bar is depicted in white color.

### 4.5 Comparative study of '$\alpha_t$' and '$\delta$' for QPI on thin and thick structures

PSI has been extensively used for the QPI of various biological cells/tissues such as RBCs, HeLa cells, macrophages and cancer tissues in the past [38-41]. However, the influence of either phase shift error '$\alpha_t$' or wrong '$\delta$' on their reconstructed phase maps has not been taken care of previously. The modulated background phase error (~ 100 – 200 mrad.) might not be crucial for the phase objects having height 2 – 4 μm (or $\varphi \sim$ 2 – 10 rad.), such as RBCs. Although, it will introduce some sort of measurement error in their reconstructed phase maps. Whereas, such modulated phase error can be a big problem for the specimens having small optical thickness (2 – 100 nm) such as shallow rib waveguide as discussed in this work or other thin cell membrane such as nanoscale holes present in the membrane of liver sinusoidal endothelial cells (LSECs) [42], tail of sperm cells [43] and *E. coli* bacteria [44] etc.

We carried out a systematic study on two structures having different thicknesses to envisage the influence of background modulation. First structure having relatively large thickness, i.e. human RBCs (thickness of around ~2.5 micron) and second structure having small thickness, i.e. optical waveguide with rib height of only 8 nm are utilized during experiments. Figure 10a and 10d depict the reconstructed phase maps of human RBCs and rib waveguide, respectively, without modulation error. It can be clearly visualized from Figs. 10b and 10c that the background modulation generated due to phase shift error '$\alpha_t$' of 12° and utilization of wrong phase shift value '$\delta$' equal to 74° instead of its actual value (i.e., 86°) is not visible with RBCs phase map. Whereas, it is found to be indispensable for 8 nm rib waveguide and affects the results significantly as presented in Figs. 10e and 10f. It is worth noting that the background modulation error generated due to $\alpha_t = 12°$ and phase shift value '$\delta$' equal to 72° rather than its actual value (i.e., 84°) greatly influence the reconstructed phase map of 8nm rib waveguide.

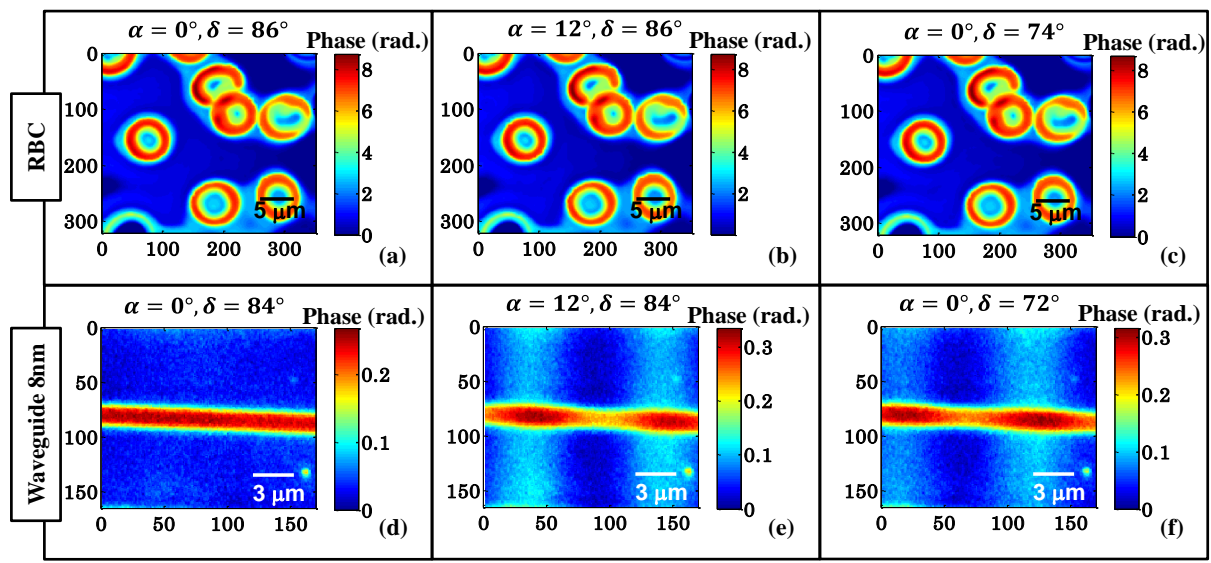

**Figure 10** Investigation of the influence of phase shift error '$\alpha_t = 12°$' and wrong '$\delta$' on QPI of human RBCs and 8nm rib waveguide. A small error in $\alpha_t$ and δ does not influence the phase images of RBC (2.5

µm thickness), while it generates strong modulation in the phase images of rib waveguide having thickness of 8 nm.

## Conclusion

Here, we demonstrate that a slight phase shift error ($\alpha_t$) introduces an unwanted background modulation in the reconstructed phase map. Similar type of modulation is also observed in the reconstruction when wrong value of phase shift 'δ' is utilized in Eq. 6. Experimentally, there are various possible reasons behind the phase shift errors such as hysteresis and nonlinearity of PZT, environmental instability and unwanted vibrations in the set-up. In order to avoid these problems, a time lapsed interferometric movie of temporally phase shifted frames is recorded. The interferometric movie contains all equal and unequal phase shifted interferograms. However, phase shift between all the data frames are unknown. A new numerical algorithm is proposed that can calculate the amount of phase shift between all frames of movie with high precision ≤ $5.5 \times 10^{-4}$ π rad. Five equally phase shifted images from the movie is then extracted to generate uncompromised quantitative phase image of the samples. Using the proposed methodology, the phase sensitivity is found to be minimal (< 1nm) while using 12-bit CMOS camera for interferometric recording. The phase sensitivity/accuracy of the PSI techniques can be further improved with the utilization of 16-bit super-cooled camera. We also demonstrated the influence of detector's noise on the phase measurement sensitivity which becomes prominent on specimen having small optical thickness (< 10nm).

The proposed way of recording temporal phase shifted interferograms and phase shift calculation algorithm can be applied under the influence of vibration/ air turbulence. The accuracy of the phase map reconstruction is found to be in close agreement to that of well calibrated PZT. The proposed approach has the capability to enable many unconventional methods like translation of reference or sample arm manually, natural vibrations/air turbulence, cell phone vibration, and artificial air turbulence generated from hair dryer etc. for introducing temporal phase shift between data frames.

The PSI has mostly implemented for the phase imaging of thick biological/industrial samples such as RBCs, HeLa cells, macrophages and cancer tissues previously, where minute phase variation will not be visible [38-41]. Therefore, background modulation (100 – 200 mrad.) would be difficult to see along with thick sample's phase map (2 – 4 rad.). Here, we demonstrated that a slight phase shift error (such as '$\alpha_t = 12°$') or utilization of wrong 'δ' is crucial in PSI for the accurate phase/height measurements of objects having sub-10 nm height. The origin and the complete removal of these modulated background error on objects having small thickness (<10 nm) were unaddressed previously. By using the proposed methodology we successfully measured 2 nm rib height of an optical waveguide with a background noise of 4 Å. Interestingly these measurements were done without using any PZT, in presence of ambient environmental fluctuations.

The proposed approach can be implemented for optical testing of the surface profiles of large optics (use in space applications) in a hostile environment inside thermal vacuum chambers, where severe shock vibrations (displacements ~1000 µm) appear due to cryogenic pump [45]. The present method has great application in phase shifting interferometric based optical metrology, digital holography of industrial objects and in QPI of biological cells and tissues without any priori calibration of phase shifts. Thus the technology can be implemented in a robust environment without using any vibration isolation optical tables and PZT.

The time requirement for obtaining five equal phase shifted frames from set of large number of frames (e.g. 200) can be reduced by using high-speed camera. For instance a high speed camera operating at 1000 fps will record 200 frames in only 0.05-0.2 second. In the present work, the technique is implemented only for the stationary objects. The present technique can be further extended to the phase imaging of moving objects with the incorporation of subpixel shift image registration algorithm [46]. Therefore, the proposed method can also be applied to generate phase images with extremely high accuracy of fast moving living cells or sub-cellular structures [1, 3, 47].

**Acknowledgement:**

D.S.M acknowledges Department of Atomic Energy (DAE), Board of Research in Nuclear Sciences (BRNS) for financial grant no. 34/14/07/BRNS. B.S.A acknowledges the funding from the European Research Council, (project number 336716) and Norwegian Centre for International Cooperation in Education, SIU-Norway (Project number INCP- 2014/10024).

**Author contributions**

Nature Communications requires an author contributions statement as described in the Authorship section of our joint Editorial policies.

**Competing interests**

A competing interests statement is required for all content of the journal. This statement will be published at the end of all papers, whether or not a competing interest is reported.

**Supplementary Note for Ahmad et al., "." :**

**Visualization 1:** Time lapsed movie of the five phase shifted interferograms of a strip waveguide (Height ~ 220nm) using uncalibrated PZT.

**Visualization 2:** The interferometric movie of the temporal phase shifted data frames introduced by hair dryer.

**Visualization 3:** The interferometric movie of the temporal phase shifted data frames of strip waveguide by translating reference or sample arm manually.

**Visualization 4:** Time lapsed interferometric movie of human RBCs with a continuous phase shift between frames introduced manually.

**Visualizations 5 – 7:** Time lapsed interferometric movie of 8nm rib waveguide recorded by 8-bit color CCD Infinity2-1RC, 8-bit monochrome CCD and 12-bit monochrome ORCA-spark C11440-36U CMOS cameras respectively. The continuous temporal phase shift between frames is introduced manually.

**Visualization 8:** Time lapsed interferometric movie of 2nm rib waveguide recorded by 12-bit monochrome ORCA-spark C11440-36U CMOS camera. The continuous temporal phase shift between frames is introduced manually.

**Additional information:**

**Simulated five equal phase shifted interferograms:**

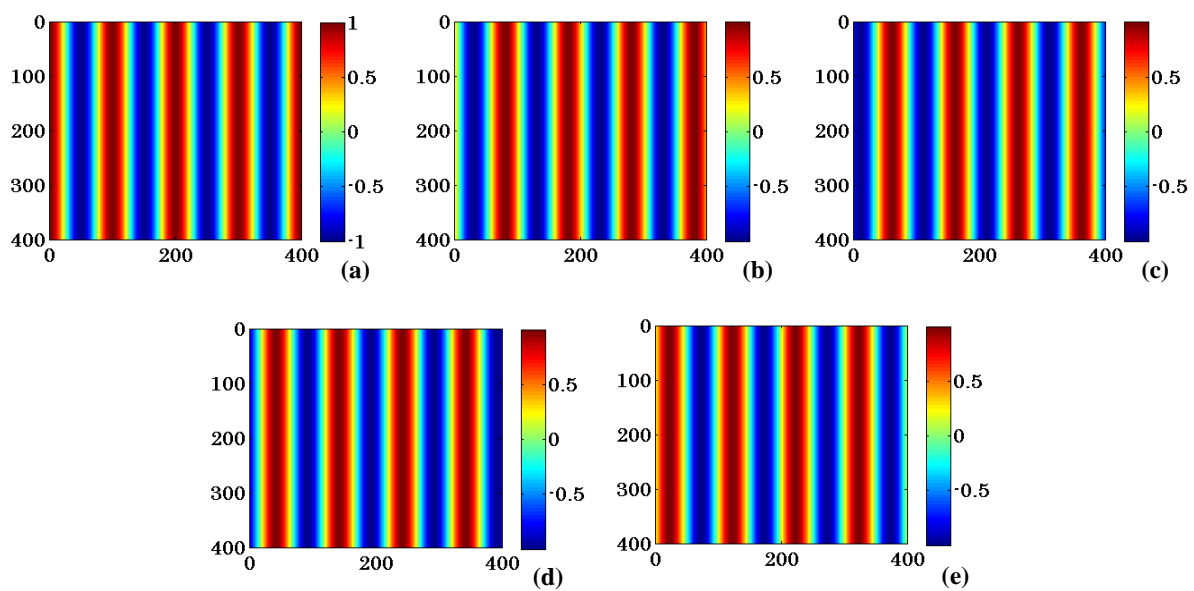

**Supplementary Fig. S1.** Simulated five equal phase shifted interferograms having phase shift ~ 70° between data frames.

**Influence of phase shift error 'α' on the reconstructed phase map:**

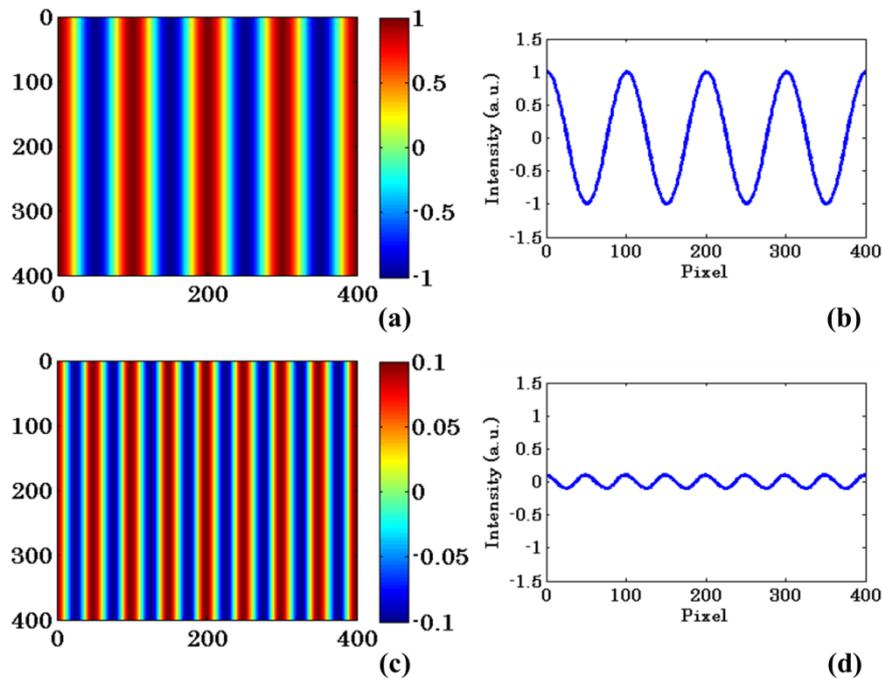

**Supplementary Fig. S2.** Effect of phase shift error 'α' between data frames in the reconstructed phase map. (a) Simulated interferogram and (b) corresponding line profile along 200$^{th}$ row of the interferogram. (c) Reconstructed phase map by employing five frame phase shifting algorithm having phase shift error 'α' equal to 20° in one of the five phase shifted data frames, and (d) corresponding line profile along 200$^{th}$ row of the reconstructed phase map.

**Experimentally recorded five equal phase shifted interferograms using well calibrated PZT:**

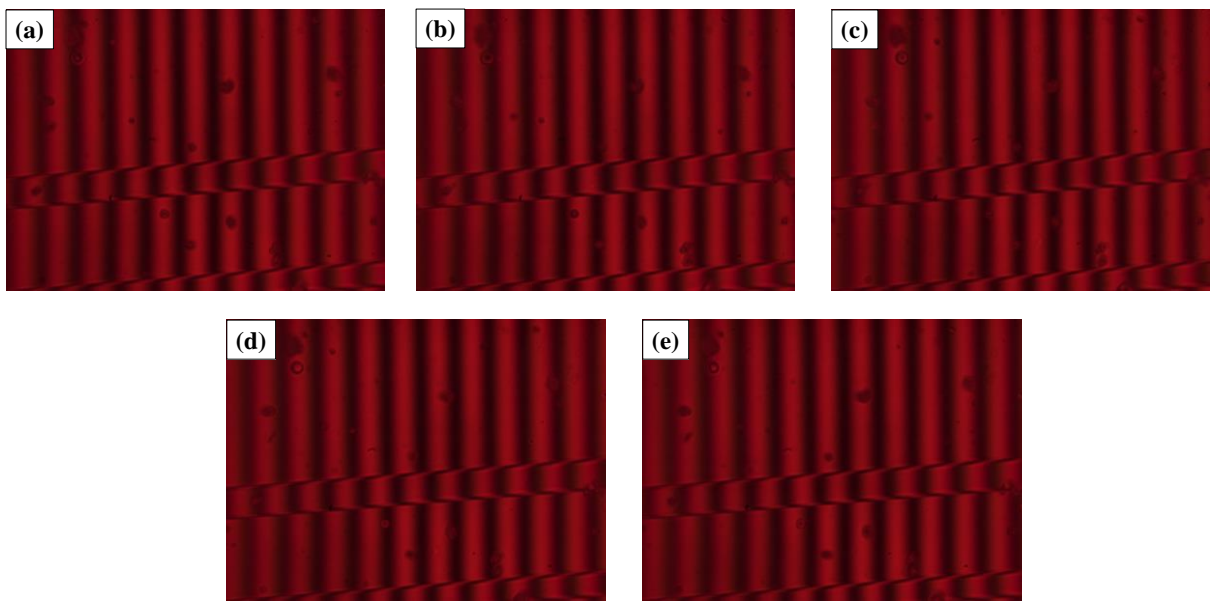

**Supplementary Fig. S3.** Experimentally recorded five equal phase shifted interferograms using calibrated PZT having phase shift ~ 75° between data frames.

**Spectral response of three different CCD/CMOS camera:**

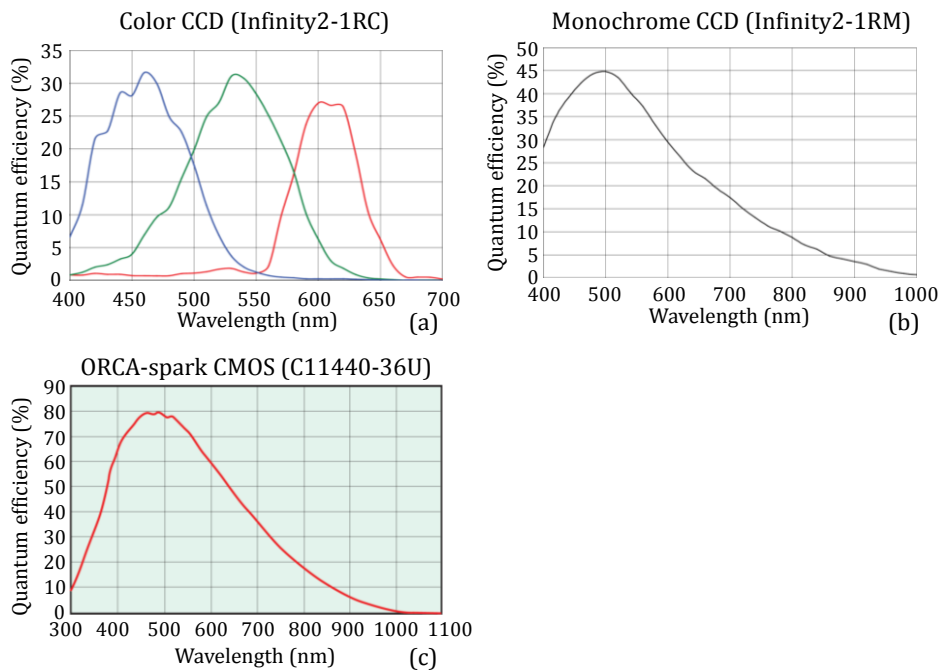

**Supplementary Fig. S4.** Spectral response of three different recording devices: (a) Color CCD (Infinity2-1RC), (b) Monochrome CCD (Infinity2-1RM), and (c) Monochrome CMOS (C11440-36U) camera [37, 48].

**Color cross talk in Color CCD (Infinity2-1RC) camera:**

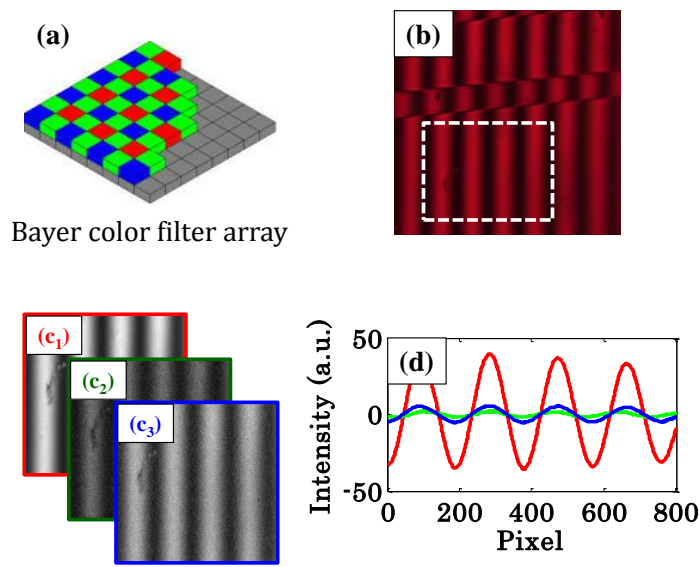

**Supplementary Fig. S5.** Experimental investigation of color crosstalk of color CCD camera when a bandpass color filter having 620nm central wavelength and ~40nm bandwidth into the white light beam path. (a) Bayer color filter array used in color CCD camera for recording red, green and blue wavelength information in different color channels of the camera. (b) Color interferogram recorded using color CCD camera. ($c_1$ – $c_2$) color decomposed interferograms using MATLAB. (d) Corresponding line profiles along middle row of the color decomposed interferograms.

**Important parameters of the recording devices for phase measurement:**

**Supplementary Table S1.** Specifications of three scientific cameras which play a vital role in quantitative phase measurements [37, 48].

| Camera Specifications | Color CCD (Infinity2-1RC) | Monochrome CCD (Infinity2-1RM) | Monochrome CMOS (C11440-36U) |
|---|---|---|---|
| Pixel Size | $4.65 \times 4.65$ μm | $4.65 \times 4.65$ μm | $5.86 \times 5.86$ μm |
| Bit Depth | 8-bit | 8-bit | 12-bit |
| Dynamic Range | 64.6 dB | 64.6 dB | 74 dB |
| Full Well Capacity | 14,500 e$^-$ | 14,500 e$^-$ | 33,000 e$^-$ |
| Quantum Efficiency (%) | 32% | 44% | 80% |
| Read Noise | 8.5 e$^-$ | 8.5 e$^-$ | 6.6 e$^-$ |

**Five phase shifted frames:**

**Supplementary Table S2.** Five equal phase shifted interferometric frames of the recorded movies obtained from three different cameras:

| S. No. | Camera | Interferometric movie's frame number | Phase shift 'δ' between consecutive frames in degree |
|---|---|---|---|
| 1. | Infinity2-1RC | 1, 54, 87, 129, and 166 | 72.6° |
| 2. | Infinity2-1RM | 1, 104, 147, 224, and 259 | 70.2° |
| 3. | C11440-36U | 1, 56, 89, 134, and 171 | 84.9° |